\documentclass[prd,twocolumn,amsmath,amssymb,floatfix,superscriptaddress]{revtex4}

\usepackage{graphicx}
\usepackage{amssymb}
\usepackage{amsmath}
\usepackage{color}
\usepackage{ wasysym }
\usepackage{hyperref}
\usepackage{tabu}
\usepackage{pdflscape}
\usepackage{bm}
\usepackage{hyperref}
\usepackage{nameref}
\def\barray{\begin{array}}
\def\earray{\end{array}}
\def\be{\begin{equation}}
\def\ee{\end{equation}}
\def\ben{\begin{equation} \nonumber}
\def\een{\end{equation}}
\def\ban{\begin{eqnarray*}}
\def\ean{\end{eqnarray*}}
\def\ba{\begin{eqnarray}}
\def\ea{\end{eqnarray}}

\def\({\left(}
\def\){\right)}

\graphicspath{{./fig/}}

\begin{document}

\title{$\Lambda$CDM Model Against Gravity-Thermodynamics Conjecture: Observational Constraints After DESI 2024}
\author{N. Shahhoseini}
%\email{malekjani@basu.ac.ir}
\affiliation{Department of Physics, Bu-Ali Sina University, Hamedan
	65178, 016016, Iran}
\author{M. Malekjani}
\email{malekjani@basu.ac.ir }
\affiliation{Department of Physics, Bu-Ali Sina University, Hamedan
	65178, 016016, Iran}
\author{A. Khodam-Mohammadi}
%\email{s.pouri90@gmail.com}
\affiliation{Department of Physics, Bu-Ali Sina University, Hamedan
	65178, 016016, Iran}
%\author{N. Shahhoseini, M. Malekjani and A. Khodam-Mohammadi}

%\thankstext{e}{e-mail: malekjani@basu.ac.ir}
%\institute{Department of Physics, Bu-Ali Sina University, Hamedan 65178, 016016, Iran.}
\date{}

\begin{abstract}
In the context of the gravity-thermodynamics conjecture based on the Tsallis entropy-area relation, we investigate the standard $\Lambda$CDM cosmology and examine some potential deviations from it. Utilizing recent updates from geometrical datasets, including the DESI BAO measurements (2024), Planck CMB anisotropy measurements (2018), and the Pantheon+ catalogue for SNIa (2022), we conduct a thorough analysis via the window of Tsallis cosmology. In the first step, our analysis reveals no significant deviation from the standard model when using DESI BAO data alone, Pantheon+ data alone, or a combination of both. In the next step, we incorporate all datasets by adding the CMB data to our analysis, indicating a potential deviation from the standard model within the framework of Tsallis cosmology. Imposing the Planck prior to the sound horizon at the baryon drag epoch, we observe support for the standard model and consistency between our constraints on the Hubble constant and the Planck value. Finally, we compare the Tsallis and $\Lambda$CDM cosmologies using the Akaike Information Criterion (AIC).
    
\end{abstract}
\maketitle

\section{Introduction}
Numerous independent cosmological observations have confirmed the accelerated expansion of the Universe. This phenomenon is evidenced by data from Type Ia supernovae (SNIa), which act as 'standard candles' for measuring cosmic distances \citep{Riess:1998cb,Perlmutter:1998np,Kowalski:2008ez}. Additionally, the cosmic microwave background (CMB) provides a snapshot of the early Universe, offering insights into its subsequent evolution \citep{Komatsu2009,Jarosik:2010iu,Ade:2015rim}. Weak gravitational lensing \citep{Benjamin:2007ys,Amendola:2007rr,Fu:2007qq}, baryon acoustic oscillations (BAO), and the large-scale structure of the Universe \citep{Tegmark:2003ud,Cole:2005sx,Eisenstein:2005su,Percival2010,Blake:2011en,Reid:2012sw} further support this accelerated expansion. Moreover, distributions of high-redshift galaxies and galaxy clusters provide additional evidence for this phase of accelerated expansion \citep{Bunker_2010}.\\
Within the framework of general relativity (GR), the observed cosmic acceleration is attributed to a mysterious component with negative pressure, commonly referred to as dark energy (DE). A widely accepted theory suggests that DE consists of vacuum energy, also known as the Cosmological Constant $\Lambda$, characterized by a constant equation of state (EoS) parameter $w_{\Lambda} = -1$ \citep{Peebles:2002gy}. This Cosmological Constant, along with Cold Dark Matter (CDM), forms the foundation of the standard cosmological model, known as the $\Lambda$CDM model. Theoretically, the $\Lambda$CDM framework faces challenges such as the 'fine-tuning' and 'cosmic coincidence' problems \citep{Weinberg:1988cp,Sahni:1999gb,Carroll:2000fy,Padmanabhan:2002ji}. Observationally, it encounters issues like the 'Hubble tension' and the '$S_8$ tension', which question the model's completeness within the standard $\Lambda$-cosmology. For a recent review addressing the observational tensions within the $\Lambda$CDM cosmology, see \citep{Perivolaropoulos:2021jda}.\\
The ongoing challenges within the standard cosmological model have driven cosmologists to investigate alternatives that extend beyond the conventional $\Lambda$CDM cosmology. Two primary approaches are being actively explored: one suggests a dynamic energy density for DE with significant negative pressure, aligning with the principles of general relativity (GR); the other advocates for a fundamental revision of GR itself, through the lens of modified gravity (MG) theories (For a review, see \citep{Tsujikawa:2010zza}). Additionally, gravitational thermodynamics has emerged as a compelling method for examining the accelerating expansion of the universe through the modification of entropy-area relation \citep{Cai:2005ra}. This approach stems from the foundational concepts of black hole physics developed by Hawking and Bekenstein \citep{Hawking:1975vcx,Bekenstein:1973ur}. They demonstrated that black holes can be treated as thermodynamic systems that adhere to the laws of thermodynamics. Consequently, it is possible to assign quantities such as temperature and entropy to their event horizons. The temperature is proportional to the surface gravity, $\kappa$, via the relation $T_H = \kappa / 2\pi$, and the entropy is proportional to the surface area, $A$, of the event horizon via $S_{BH} = A / 4G$.
Entropy and temperature are recognized as thermodynamic quantities, while surface area and surface gravity are geometric quantities. Since the discovery of black hole thermodynamics and the realization of the proportionality between these quantities, it has been suggested that the fundamental equations governing these two theories may correspond. Jacobson \citep{Jacobson:1995ab} was the first to explore the connection between the laws of thermodynamics and Einstein's field equations. He derived Einstein's field equations by applying the fundamental Clausius relation, $\delta Q = T dS$, and treating the black hole horizon as the system's boundary. Subsequently, numerous efforts have been made to investigate the deeper association between gravity and thermodynamics, demonstrating that this connection exists in various other gravitational theories as well \citep{Eling:2006aw,Akbar:2006er,Akbar:2006mq,Padmanabhan:2003gd,Padmanabhan:2009vy,Padmanabhan:2002sha,Padmanabhan:2006fn,Paranjape:2006ca,Kothawala:2007em,Padmanabhan:2007en}. The idea of these investigations has been entered to cosmological contexts, where thermodynamic laws have been applied to the apparent horizon in cosmology. This application has lead to the derivation of the Friedmann equations from the first law of thermodynamics \citep{Akbar:2006kj,Cai:2006rs,Sheykhi:2007zp,Sheykhi:2007gi,Jamil:2009eb,Cai:2009ph,Wang:2009zv,Jamil:2010di,Gim:2014nba,Fan:2014ala,DAgostino:2019wko,Sanchez:2022xfh}.
 %However, in some gravitational theories, this area law must be modified accordingly \citep{Wald:1993nt}.
 While applying the Bekenstein-Hawking entropy relation to the apparent horizon in cosmology yields the standard Friedman equation, this relation is typically used for extensive systems \citep{Biro:2011ncf,Biro:2013cra,Komatsu:2013qia,Majhi:2017zao}. However, the universe as a whole may be considered a non-extensive system. In this context, several non-extensive entropies, such as Tsallis \citep{Tsallis:1987eu}, Barrow \citep{Barrow:2020tzx}, and Kaniadakis \cite{Lymperis:2021qty}, have been recently proposed. These proposals, which modify the entropy-area relation, lead to modified Friedman equations that extend beyond the standard $\Lambda$CDM cosmology. 
 Through the window of the modified entropy-area relation, numerous efforts have been made to address and mitigate key issues in cosmology. These include inflationary cosmology \cite{Khodam-Mohammadi:2024iuo, Odintsov_2023, Odintsov:2023rqf, Lambiase:2023ryq, Teimoori:2023hpv, Luciano:2023roh}, the positive acceleration of the expanding Universe in dark energy scenarios \cite{Brevik:2024nzf, Sheykhi_2023, P:2022amn, Nojiri_2022, Manoharan:2022qll, Di_Gennaro_2022, Bhattacharjee_2021, Moradpour:2020dfm, DAgostino:2019wko, Saridakis_2018}, and addressing or alleviating the Hubble constant ($H_0$) tension and the amplitude of matter fluctuation ($\sigma_8$) tension \cite{Basilakos:2023kvk, Hern_ndez_Almada_2022, Asghari:2021bqa, Dabrowski:2020atl}. Notably, within the framework of non-extensive entropy scenarios, the modified Friedman equations can be expressed to include new terms representing the effective dark energy component. For details on Tsallis entropy, see \cite{Lymperis:2018iuz}.\\
While numerous theoretical proposals extend beyond the standard model, validating these theories using available cosmological data is crucial. Investigating potential deviations from the standard $\Lambda$CDM model through the window of dynamical DE \cite{DESI:2024mwx,Pourojaghi:2024tmw} and modifications of gravity \cite{Chudaykin:2024go} can be assessed using cosmological data across various scales and redshifts.
In this study, we utilize the latest major observational data, including the first-year release of the Dark Energy Spectroscopic Instrument (DESI) for BAO measurements in 2024 \cite{DESI:2024mwx}, CMB Planck anisotropy observations in 2018 \cite{Planck:2018vyg}, and the Pantheon+ sample for SNIa in 2021 \cite{Scolnic:2021amr}. We place cosmological constraints on the Tsallis  cosmologies, addressing whether this model deviates from the standard $\Lambda$CDM cosmology from the perspective of cosmological observations.

The paper is organized as follows: In Sect. \ref{Sect:2}, we introduce the general thermodynamical aspects of gravitation. In Sects. \ref{Sect:3}, the modified FLRW Universe is introduced in the context of Tsallis entropy. In Sect. \ref{Sect:5}, we introduce different combinations of the cosmological data used in our analysis and present our numerical results.
In Sect. \ref{Sect:6}, we conclude our study.

\section{Thermodynamical aspects}\label{Sect:2}
In a homogeneous and isotropic FLRW spacetime,
\begin{equation}
ds^2=h_{ab}dx^a dx^b+R^2(d\theta^2+sin^2\theta d\phi^2),
\end{equation}
where $x^i=(t,r),~ R=a(t)r$ and $h_{a b}=diag\Big{(}-1,~a(t)^2/(1-kr^2)\Big{)}$ with its determinant, `$h$'. To examine the universe as a thermodynamic system, we need to consider a thermodynamical horizon of the universe, apparent horizon, defined by $h^{ab}\partial R_a \partial R_b=0$ \cite{Sanchez:2022xfh,Cai:2005ra,Binetruy:2014ela,Hayward:1997jp}. In a FLRW universe, its solution yields the apparent horizon with the radius 
\begin{eqnarray}\label{eq19}
r_A=\frac{1}{\sqrt{H^{2}+\frac{k}{a^{2}}}}.
\end{eqnarray}
Therefore, the area of the apparent horizon and the volume enclosed by a three-dimensional sphere are defined as follows
\begin{eqnarray}
A&=&4\pi r_A^{2}\label{eq20}\\
V&=&\frac{4}{3}\pi r_A^{3}.\label{eq21}
\end{eqnarray}
The first law of thermodynamics within the boundary of the apparent horizon is 
\begin{eqnarray}\label{eq22}
dE=TdS+WdV,
\end{eqnarray}
where $S$ is the gravitational entropy and the work density $W=-1/2 T^{a b}h_{a b}$, which becomes from a perfect fluid energy-momentum density $T^\mu_\nu=diag\Big{(}-\rho,p,p,p\Big{)}$
\begin{eqnarray}\label{eq24}
W=\frac{1}{2}(\rho - p).
\end{eqnarray}
The temperature $T$ is obtained by the surface gravity $\kappa$ as \citep{Sheykhi:2013oac,Cai:2005ra,Akbar:2006kj}
\begin{eqnarray}\label{eq25}
\kappa&=&\frac{1}{2\sqrt{-h}}\frac{\partial}{\partial x^a}(\sqrt{-h}h^{a b}\frac{\partial r_A}{\partial x^b}),\\
T_h&=&\frac{|\kappa|}{2\pi}=\frac{1}{2\pi r_A}|1-\frac{\dot{r_A}}{2 H r_A}|.\label{eq23}
\end{eqnarray}
The conservation of energy-momentum  $T^{\mu\nu}_{;\mu}=0$ leads the continuity equation
\begin{eqnarray}\label{conteq}
\dot{\rho}+3H(\rho+p)=0.
\end{eqnarray}
In the case of a pure de Sitter space where, $\rho=-p$ , the work term, $WdV$, simplifies back to  $-pdV$ in the standard form of the first law of thermodynamics, $dE=TdS-pdV$.

Currently, our primary focus is to consider the role of entropy in FLRW cosmology. Our objective is to formulate the Friedmann equations considering the non-extensive entropies. Hence, in the subsequent sections, we will delve into this subject comprehensively and work towards formulating the Friedmann equations as well as examining the significant dimensionless Hubble parameter.

\section{Modification of FLRW cosmology through Tsallis  entropy}\label{Sect:3}
In 1902, Gibbs pointed out that the conventional Boltzmann-Gibbs framework is inadequate for systems with a non-converging partition function, particularly for large-scale gravitational systems. Consequently, the standard additive entropy of Boltzmann Gibbs, where the total entropy of a system is merely the sum of its parts, must be generalized to a non-additive form. This implies that the entropy of the entire system is not simply the sum of the entropies of its subsystems \citep{Tsallis:1987eu,Lyra:1997ggy,Wilk:1999dr,Sheykhi:2019bsh}. In 1988, Tsallis expanded the realm of traditional thermodynamics to include non-extensive thermodynamics, which is universally applicable while still encompassing the classical Boltzmann-Gibbs theory as a special case \citep{Tsallis:1987eu}. Building on this foundation, Tsallis and Cirto \citep{Lyra:1997ggy} used statistical arguments to propose that the entropy of a black hole does not adhere to the area law and should be modified accordingly:
\begin{eqnarray}\label{eq26}
S=\gamma A^{\beta}.
\end{eqnarray}
In cosmology, $A$ denotes the apparent horizon area of the universe, $\gamma$ is a constant, and $\beta$ is referred to as the Tsallis parameter or the non-extensive parameter. When $\beta=1.0$ and $\gamma=\frac{1}{4G}$, the standard entropy-area relation (Bekenstein-Hawking entropy) is recovered. \citep{Jizba:2022bfz}.
Differentiating the modified entropy-area relation given in equation(\ref{eq26}), we obtain
\begin{eqnarray}\label{eq27}
	dS=8 \pi \gamma \beta (4\pi r_{A}^{2})^{\beta -1} r_{A} dr_{A}
\end{eqnarray}
To derive the first Friedmann equation within the framework of Tsallis cosmology, we utilize the relations (\ref{eq24}), (\ref{eq25}), (\ref{eq23}), (\ref{eq27}), and the first law of thermodynamics (\ref{eq22}). Additionally, we incorporate the continuity equation (\ref{conteq}). By applying the apparent horizon relation (\ref{eq19}) and performing some straightforward calculations, we obtain the first modified Friedmann equation based on Tsallis entropy as follows \cite{Sheykhi:2018dpn}:
\begin{eqnarray}\label{eq28}
	(H^{2}+\frac{k}{a^{2}})^{2-\beta} = \frac{8 \pi G}{3} \rho_{t}
\end{eqnarray}
where $\rho_{t}=\rho_m + \rho_r + \rho_{\Lambda}$ is the sum of all energy densities of the various components of the universe including pressureless matter $\rho_m$, radiation $\rho_r$ and cosmological constant $\rho_{\Lambda}$.
The second Friedman equation can be obtained by taking the time derivative of equation (\ref{eq28}) and using the continuity equation (\ref{conteq}), as follows \citep{Sheykhi:2019bsh,Sheykhi:2018dpn,Sheykhi:2022gzb}.
\begin{eqnarray}
\nonumber
(4-2\beta) \frac{\ddot a}{a}(H^{2}+\frac{k}{a^{2}})^{1-\beta}+ (2\beta-1)(H^{2}+\frac{k}{a^{2}})^{2-\beta}
\\
=-8\pi G p.\label{eq30}
\end{eqnarray}
We can obtain  the following relation for $\gamma$,
\begin{eqnarray}\label{eq29}
	\gamma \equiv \frac{2-\beta}{4 \beta G} (4\pi)^{1-\beta},
\end{eqnarray}
where putting $\beta=1.0$, leading $\gamma=\frac{1}{4G}$ in standard Bekenstein-Hawking entropy.
Eq.\ref{eq30} represents the altered version of the second Friedmann equation, which dictates the dynamics of the Universe's expansion within the framework of Tsallis cosmology. Notably, for $\beta=1.0$, the conventional second Friedmann equation is recovered.
In this work, we consider a non-interacting and spatially flat FLRW Tsallis cosmology. In this context, the energy density of matter evolves as $\rho_m=\rho_{m,0} (1+z)^3$, radiation component evolves as $\rho_r=\rho_{r,0} (1+z)^4$, and for any '$i$' component of energy content, $\Omega_{i,0}=\rho_{i,0}/\rho_{crit,0}$. Notice that the recent paper by DESI collaboration \cite{DESI:2024mwx} supports the flat FLRW  universe \citep[see also][]{Efstathiou:2020wem}. In Tsallis cosmology by choosing the critical energy density as
\begin{eqnarray}\label{crittsal}
\rho_{crit,0}=\frac{3H_0^{4-2\beta}}{8\pi G},
\end{eqnarray}
the normalized Hubble parameter can be obtained from (\ref{eq28}) as follows 
\begin{eqnarray}\label{eq31}
&&E(z)=\frac{H(z)}{H_0}\nonumber\\
&=&\left[\Omega_{m,0} (1+z)^{3}+\Omega_{r0} (1+z)^{4} + \Omega_{\Lambda} \right]^{\frac{1}{4-2\beta}},
\end{eqnarray}
where in the flat FLRW universe $\Omega_{\Lambda} = 1-\Omega_{m,0} - \Omega_{ro}$.
Clearly, setting $\beta=1.0$ restores the standard flat $\Lambda$CDM cosmology. Any deviation from $\beta=1.0$ suggests a potential deviation from the standard cosmological model.
\section{Observational datasets and cosmological constraints}\label{Sect:5}
In this section, we utilize various combinations of the main observational datasets from both high and low redshifts to constrain the cosmological parameters within the frameworks of Tsallis cosmology. To achieve this, we employ the Markov Chain Monte Carlo (MCMC) algorithm to determine the best-fit values of the cosmological parameters and their uncertainties. The most recent and updated observational data used in our analysis include the DESI BAO \citep{DESI:2024mwx}, the CMB Planck anisotropy measurements \cite{Planck:2018vyg}, and the apparent magnitude of SNIa data from the Pantheon+ catalogue \cite{Scolnic:2021amr}. We begin by briefly describing each dataset and then present our numerical results for various combinations of observational data.
\subsection{Observational datasets}
\subsubsection{DESI BAO 2024}
Baryon acoustic oscillations (BAO) act as standard rulers, reflecting the early interactions between baryons and photons in a hot plasma, which are imprinted on the matter power spectrum before recombination.
BAO measurements are based on the sound horizon at the baryon drag epoch, $r_d$. 
The recent first data release from DESI has provided 12 new BAO measurements, detailed in Table I of Ref \citep{DESI:2024mwx}. These measurements include both isotropic and anisotropic BAO data within the redshift range of 0.1 to 4.2, divided into seven redshift bins.
The anisotropic BAO measurements are given as $\frac{D_M}{r_d}$ and $\frac{D_H}{r_d}$, normalized to the comoving sound horizon at the baryon drag epoch, where $D_M$ representing the comoving distance, including a measurement of the transverse BAO mode which is given by
\begin{eqnarray}\label{eq39}
	D_M(z)=\frac{c}{H_0} \int_{0}^{z} \frac{dz^{\prime}}{E(z^{\prime})},
\end{eqnarray}
and $D_H$, the Hubble distance, which represents a measurement of the radial BAO mode, described as
\begin{eqnarray}\label{eq40}
	D_H(z)=\frac{c}{H(z)}
\end{eqnarray}
The isotropic BAO measurements are expressed as $\frac{D_V}{r_d}$, where $D_V$ is the angle-averaged distance normalized to the comoving sound horizon at the drag epoch. In this case, radial and transverse BAO modes are not separated from each other, resulting in $D_V$ expressed as
\begin{eqnarray}\label{eq41}
	D_V(z)=\left(z \: D_M(z)^{2} \: D_H(z)\right)^{\frac{1}{3}}.
\end{eqnarray}
Additionally, in this dataset the correlation between $\frac{D_M}{r_d}$ and $\frac{D_V}{r_d}$ measurements is considered. It is important to highlight that in the DESI BAO segment of our analysis, we first constrain the quantities $\frac{D_M}{r_d}$, $\frac{D_H}{r_d}$, and $\frac{D_V}{r_d}$, assuming $r_d$ as a free parameter. Notice that Table I in \cite{DESI:2024mwx} provides observational data for these ratios rather than for $D_M$, $D_H$, and $D_V$ individually. We then apply the Planck-CMB measurements on $r_d$ as a physical prior to the DESI BAO segment of our analysis.

	\begin{figure} 
		\centering
		\includegraphics[width=8cm]{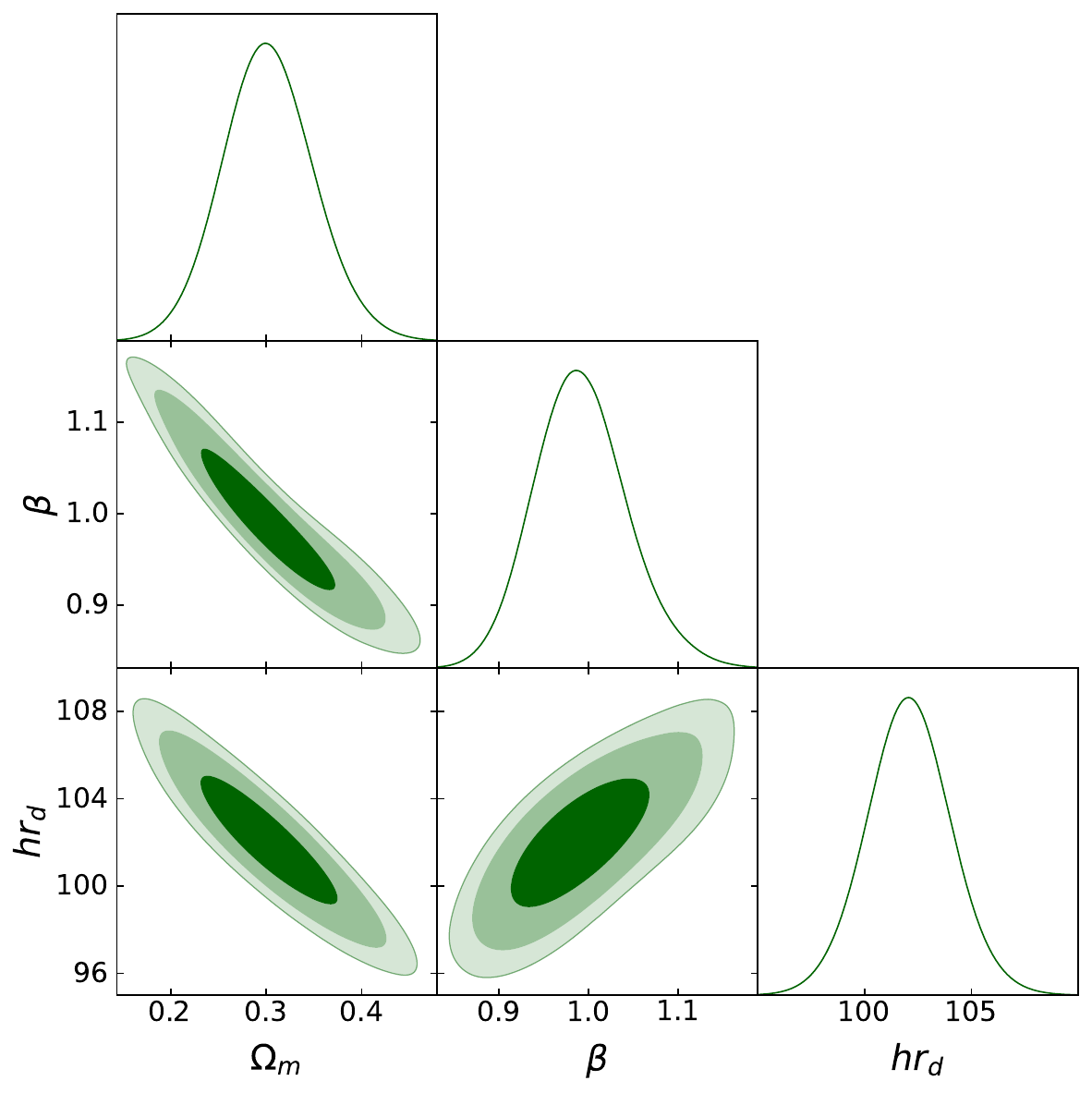}
		\caption{The confidence contours at $1\sigma$ , $2\sigma$ and $3\sigma$ levels of the cosmological parameters $\Omega_{m0}$, $\beta$ and $hr_d$in Tsallis cosmology from DESI BAO alone.}
		\label{fig:TDESI}
	\end{figure}
 
\subsubsection{Planck 2018}
The Planck Collaboration has recently published the final observations (Planck CMB 2018) on the temperature and polarization anisotropies of the CMB \cite{Planck:2018vyg}. In this work, we use the reduced Planck CMB measurements presented based on the distance priors. The new distance priors based on Planck CMB 2018, are presented in \citep{Chen:2018dbv}. Indeed, these distance priors capture the essential details of the CMB power spectrum in two main aspects: the acoustic scale,$l_A$, describes the CMB temperature power spectrum in the transverse direction, leading to changes in the spacing of the peaks; while the shift parameter,$R$, affects the CMB temperature spectrum in the line-of-sight direction, influencing the heights of the peaks. The definitions of these parameters are given by the following equations:
\begin{eqnarray}\label{eq43}
l_A = (1+z_\star) \frac{\pi D_A(z_\star)}{r_d(z_\star)}
\end{eqnarray}
\begin{eqnarray}\label{eq44}
R(z_\star) \equiv \frac{(1+z_\star) D_A(z_\star) \sqrt{\Omega_m H_0^{2}}}{c^{2}}
\end{eqnarray}
where $r_d$ represents the comoving sound horizon, defined as:
\begin{eqnarray}\label{eq:rd_cmb}
r_d = \int_{z_\star}^{\infty} \frac{c_s(z)}{H(z)} \, dz
\end{eqnarray}
and $z_\star$ denotes the redshift at the baryon drag epoch. The sound speed in the baryon-photon fluid, $c_s(z)$, is given by:
\begin{eqnarray}
c_s(z) = \frac{c}{\sqrt{3(1 + R(z))}}
\end{eqnarray}
where $R(z)$ is defined as:
\begin{eqnarray}
R(z) = \frac{3 \rho_b(z)}{4 \rho_\gamma(z)}
\end{eqnarray}
Additionally, the angular diameter distance of the Universe at the drag epoch, $D_A(z_\star)$, is given by:
\begin{eqnarray}
D_A(z_\star) = \frac{c}{1+z_\star} \int_0^{z_\star} \frac{dz}{H(z)}
\end{eqnarray}
It is important to note that, unlike in the DESI BAO segment, here $r_d$ is not a free parameter and is computed via Eq. \ref{eq:rd_cmb}.\\
\subsubsection{Pantheon+} 
Distant Type Ia Supernovae are among the most recognized and frequently employed tools in cosmology. These explosions are extraordinarily bright, often matching the luminosity of their host galaxies. Their light curves show the same peak brightness, making them dependable standard candles.
In this study, we utilize the Pantheon+  dataset, which is the latest compilation of spectroscopically confirmed Type Ia supernovae (SNIa) \cite{Scolnic:2021amr}. The Pantheon+ catalogue includes 1701 light curves, spanning a redshift range of 0.001 to 2.26. Among these, 77 supernovae are from galaxies that host Cepheids within the low redshift range of $0.00122$ to $0.01682$. 

\begin{figure} 
	\centering
	\includegraphics[width=8cm]{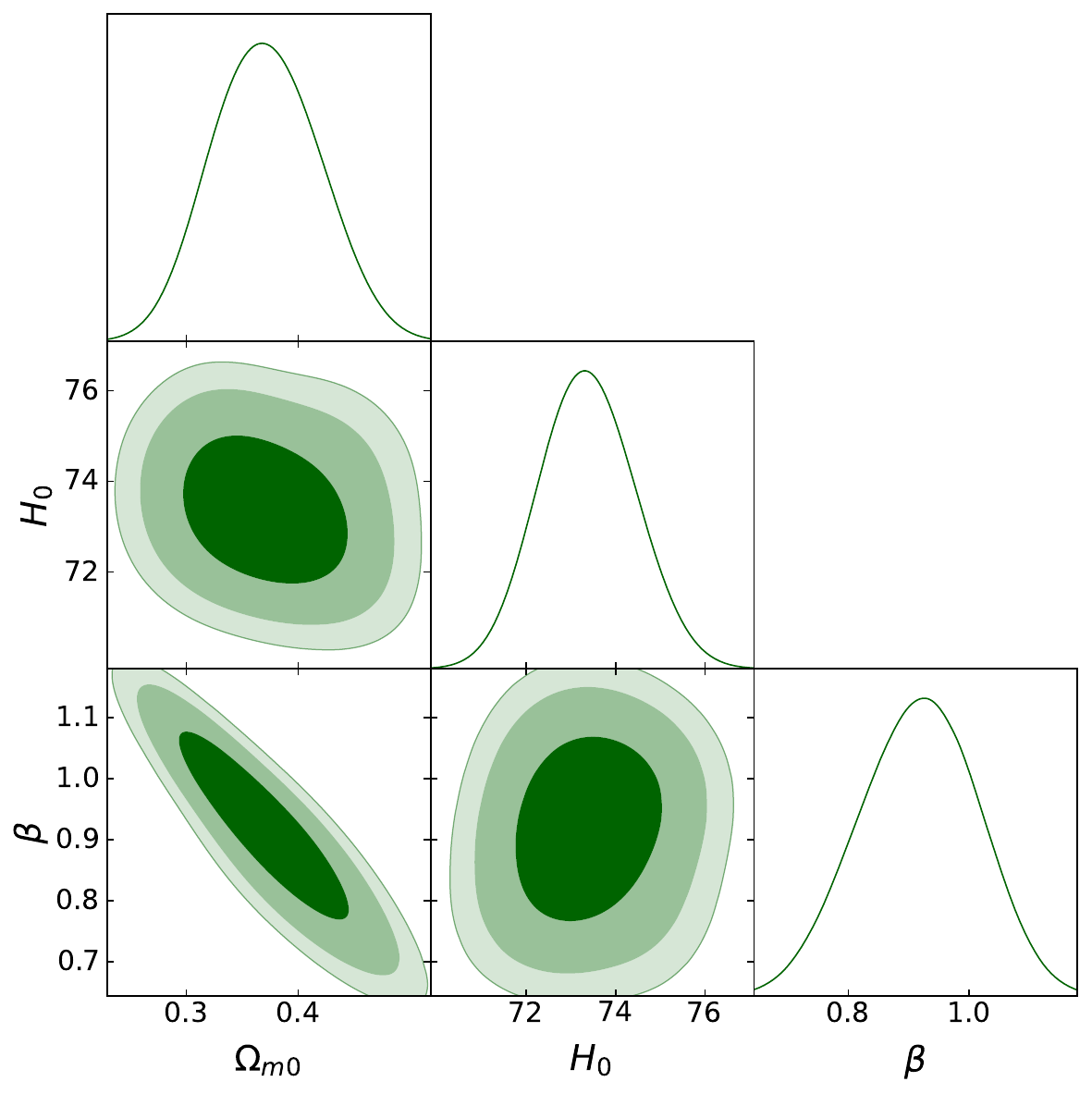}
	\caption{The confidence contours at $1\sigma$, $2\sigma$ and $3\sigma$ levels of the cosmological parameters $\Omega_{m0}$, $H_0$ and $\beta$ in Tsallis cosmology from Pantheon+ dataset.}
	\label{fig:Tpan}
\end{figure}
\subsection{\textbf{Cosmological constraints}}
We start with the DESI BAO observations and subsequently integrate additional datasets. We impose observational constraints on the cosmological parameters \{$H_0$, $\Omega_{m0}$, $\beta$\} within the framework of Tsallis cosmology. Additionally, we consider two other physical parameters, $r_d$ and $M$, which are treated as free parameters in a statistical sense. However, we can apply the Planck prior to $r_d$ and use Cepheid data to set a prior on $M$. Our constraints on the Tsallis parameter $\beta$ can indicate a deviation from the standard $\Lambda$CDM cosmology if the deviation of $\beta$ from the value $\beta = 1.0$ is measured with more than $3\sigma$ uncertainty. For completeness, we calculate the Akaike Information Criterion (AIC), $AIC=\chi^2_{min}+2K$ where $K$ is the number of free parameters \cite{1100705}, to compare the Tsallis cosmology with the standard flat-$\Lambda$CDM universe.
In statistical analysis, the model with the lowest AIC is considered the best among all models analyzed. For other models:\\
•  $\Delta \text{AIC} \leq 2$: Substantially supported and considered equally good as the best model.\\
•  $2 < \Delta \text{AIC} \leq 4$: Some evidence against the model with the higher AIC, but the evidence is not strong.\\
•  $4 < \Delta \text{AIC} \leq 7$: Considerably less support.\\
•  $\Delta \text{AIC} > 10$: Essentially no support and strongly disfavored.
\subsubsection{DESI BAO alone}
When using DESI BAO data alone, we can only determine the product of the sound horizon and the Hubble constant, $r_dH_0$ (assuming $r_d$ is a free parameter). We cannot separately constrain the sound horizon $r_d$ and the Hubble constant $H_0$. By combining DESI BAO data with other observational probes, such as CMB and SNIa datasets, we can resolve this degeneracy.
The $\chi^2$ function for DESI BAO data includes both isotropic and anisotropic components:
\begin{eqnarray}\label{eq42}
\chi^{2}_{\text{DESI BAO}} = \chi^{2}_{\text{iso}} + \chi^{2}_{\text{aniso}}.
\end{eqnarray}
We minimize the $\chi^2$ function to find the best-fit values of the cosmological parameters and their uncertainties. The numerical results and cosmological constraints for DESI BAO analysis are presented in the first column of Table \ref{tab:tab1} and Figure \ref{fig:TDESI}. Our constraints show that DESI BAO alone fully supports $\beta=1.0$ within a $3\sigma$ confidence level, thereby recovering the standard $\Lambda$CDM model in the context of Tsallis cosmology. We obtain $\Omega_{m0}$ very close to its Planck $\Lambda$CDM value in 2018 \cite{Planck:2018vyg}. In addition, we obtain $\Delta \text{AIC} = \text{AIC}_{T} - \text{AIC}_{\Lambda} = 2.5$, where $\text{AIC}_{T}$ and $\text{AIC}_{\Lambda}$ are the AIC values for the Tsallis and $\Lambda$CDM cosmologies, respectively. This result indicates weak evidence against the Tsallis cosmology, but it is not strong enough to disfavor the model statistically.

\begin{figure} 
	\centering
	\includegraphics[width=8cm]{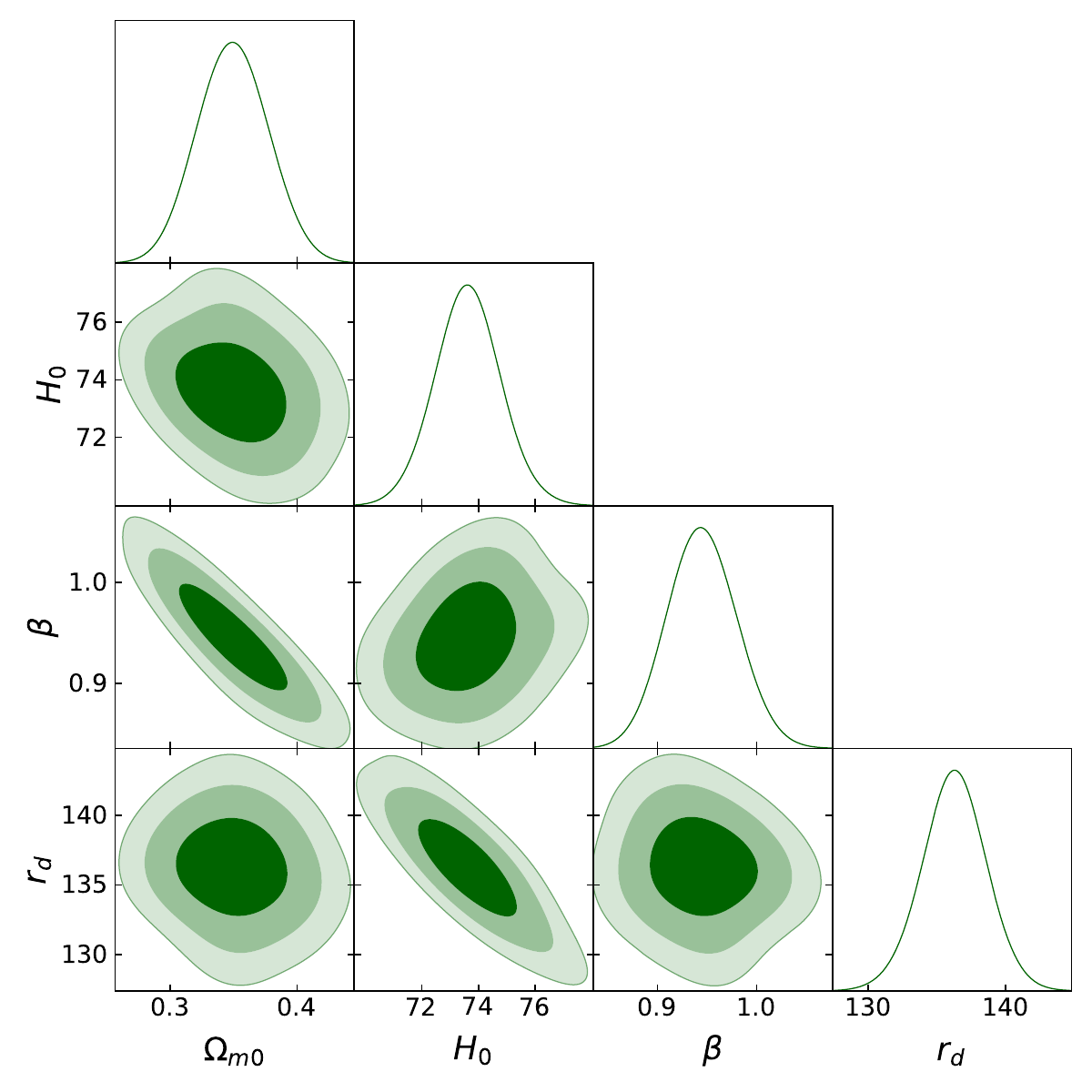}
	\caption{The confidence contours at  $1\sigma$ , $2\sigma$ and $3\sigma$ levels of the cosmological parameters $\Omega_{m0}$, $H_0$, $\beta$ and $r_d$ in Tsallis cosmology from DESI BAO + Pantheon Plus combination.}
	\label{fig:TPD}
\end{figure}

\begin{table*}
\centering
\selectfont
\caption{The Best-fit values and $3\sigma$ uncertainties for the cosmological parameters obtained in different cases of our analysis in the framework of Tsallis cosmology. For comparison, we report the $\chi^2_{min}$ and AIC values of the standard $\Lambda$CDM model in the parentheses.}
\resizebox{18cm}{2.5cm}{%
\begin{tabular}{c c c c c c }
\hline \hline
Parameter  & DESI BAO & Pantheon+ & DESI BAO + Pantheon plus & DESI BAO + Pantheon plus + Plank 2018 & Planck Prior\\
\hline 
$\Omega_{m0}$ & $0.302^{+0.045,+0.095,+0.13}_{-0.045,-0.093,-0.11}$ & $0.371^{+0.046,+0.095,+0.11}_{-0.046,-0.092,-0.095} $ & $0.349^{+0.026,+0.054,+0.071}_{-0.026,-0.050,-0.069}$ & $0.3226^{+0.0085,+0.017,+0.023}_{-0.0085,-0.018,-0.025}$ & $0.3027^{+0.0070,+0.014,+0.018}_{-0.0070,-0.013,-0.021}$\\
\hline
$H_0$ & $--$ & $73.4^{+1.0,+2.1,+2.6}_{-1.0,-1.9,-2.3}$ & $73.6^{+1.1,+2.4,+3.5}_{-1.1,-2.3,-3.0}$ & $71.75^{+0.66,+1.3,+1.8}_{-0.66,-1.3,-1.9}$ & $69.19^{+0.42,+0.87,+1.1}_{-0.42,-0.82,-1.1}$\\
\hline
$\beta$ & $0.991^{+0.042,+0.11,+0.14}_{-0.050,-0.099,-0.11} $ & $0.917^{+0.092,+0.17,+0.18}_{-0.092,-0.18,-0.19}$ & $0.945^{+0.033,+0.067,+0.10}_{-0.033,-0.065,-0.082}$ & $1.00316^{+0.00058,+0.0011,+0.0016}_{-0.00058,-0.0012,-0.0014}$ & $1.00058^{+0.00027,+0.00054,+0.00087}_{-0.00027,-0.00052,-0.00076}$\\
\hline
$r_d$ & $--$ & $--$ & $136.3^{+2.2,+4.4,+6.3}_{-2.2,-4.8,-7.0} $ & $139.1^{+1.7,+3.2,+5.0}_{-1.7,-3.3,-5.0}$ &  $147.4600^{+0.0010,+0.0020,+0.0029}_{-0.0010,-0.0020,-0.0029}$\\
\hline
$M$ & $--$ & $-19.249^{+0.029,+0.058,+0.074}_{-0.029,-0.054,-0.065}$ & $-19.245^{+0.031,+0.069,+0.10}_{-0.031,-0.066,-0.088}$ & $-19.301^{+0.020,+0.040,+0.055}_{-0.020,-0.041,-0.056}$ & $-19.385^{+0.012,+0.024,+0.033}_{-0.012,-0.023,-0.031}$\\
\hline
$hr_d$ & $102.1^{+1.9,+3.8,+5.2}_{-1.9,-3.8,-5.1} $ & $--$ & $--$ & $--$ & $--$\\
\hline
$\chi^2_{min}$ ($\chi^2_{min, \Lambda}$) & $16.3(15.7)$ & $1516.7 (1517.0)$ & $1534.7 (1535.3)$ & $1560.4 (1568.2)$ & $1563.3 (1567.8)$\\
\hline
$AIC$ ($AIC_{\Lambda}$) & $22.3(19.7)$ & $1524.7 (1523.0)$ & $1544.7 (1543.3)$ & $1570.4 (1576.2)$ & $1571.3 (1573.8)$\\
\hline\hline
\end{tabular}}\label{tab:tab1}
\end{table*}
\subsubsection{Pantheon+}
Using the Pantheon+ catalogue, we perform our MCMC analysis. The free parameters in our analysis are $\{\Omega_{m0}, H_0, M\}$, utilizing the 77 data points for the absolute magnitude of SNIa calibrated by Cepheids in the Pantheon+ catalogue \cite{Scolnic:2021amr}. Consequently, the chi-square function for the Pantheon+ dataset must account for both Cepheid and other data.
\begin{eqnarray}\label{eq45}
\chi^{2}_{Pan+}=\chi^{2}_{Cepheid} + \chi^{2}_{SNI}.
\end{eqnarray}
The results are presented in the second column of Table \ref{tab:tab1} and in Figure \ref{fig:Tpan}. Our measurements of the parameter $\beta$ indicate full support for the standard $\Lambda$CDM cosmology ($\beta=1.0$) within a $1\sigma$ confidence level, similar to what we observed for DESI BAO alone. The Pantheon+ constraints on the Hubble constant $H_0$ within the context of Tsallis cosmology are fully consistent with the SHOES value \cite{Riess:2016jrr,Riess:2020fzl,Riess:2021jrx}. Specifically, the difference between our constraint and the recent update of the SHOES value $H_0=73.04\pm1.04$ km/s/Mpc \cite{Riess:2021jrx} is around $0.25\sigma$, indicating full agreement between them. Finally, we obtain $\Delta AIC<2$, suggesting that both models have substantial support and are almost equally good.
\subsubsection{DESI BAO + Pantheon plus}
Now, we combine the DESI BAO and Pantheon+ datasets. This combination allows us to disentangle $H_0$ and $r_d$, which was not possible with DESI BAO alone. The free parameters in this analysis are $\{\Omega_{m0}, H_0, M, r_d, \beta \}$. It is important to note that we still treat $r_d$ as a free parameter to be determined by the DESI BAO observations. The results of this analysis are presented in the fourth column of Table \ref{tab:tab1} and in Figure \ref{fig:TPD}. 
We observe that the Tsallis parameter $\beta$ does not significantly deviate from the $\Lambda$CDM value $\beta=1.0$ within a $3\sigma$ statistical error. Thus, alignment of observations from DESI BAO alone and Pantheon+ alone, their combination supports the standard $\Lambda$CDM cosmology within the framework of the gravity-thermodynamics conjecture defined by modified Tsallis entropy. Additionally, we find that the constraint on the Hubble constant $H_0$ obtained from the combination of DESI BAO and Pantheon+ catalogue is in full agreement with the SHOES value \cite{Riess:2021jrx}, with a very small deviation of about $0.38\sigma$. In the next step, we compare our measurement of the sound horizon $r_d$ using the DESI BAO + Pantheon+ compilation with the Planck CMB value $r_d=147.46\pm0.28$ Mpc \cite{Lemos:2023qoy}. Notably, we find a $5.5\sigma$ deviation between our measurement and the Planck value. Statistically comparison, we get $\Delta AIC<2$, indicating that both Tsallis and $\Lambda$CDM cosmologies have substantial support and are equally good.\\
\begin{figure} 
	\centering
	\includegraphics[width=8cm]{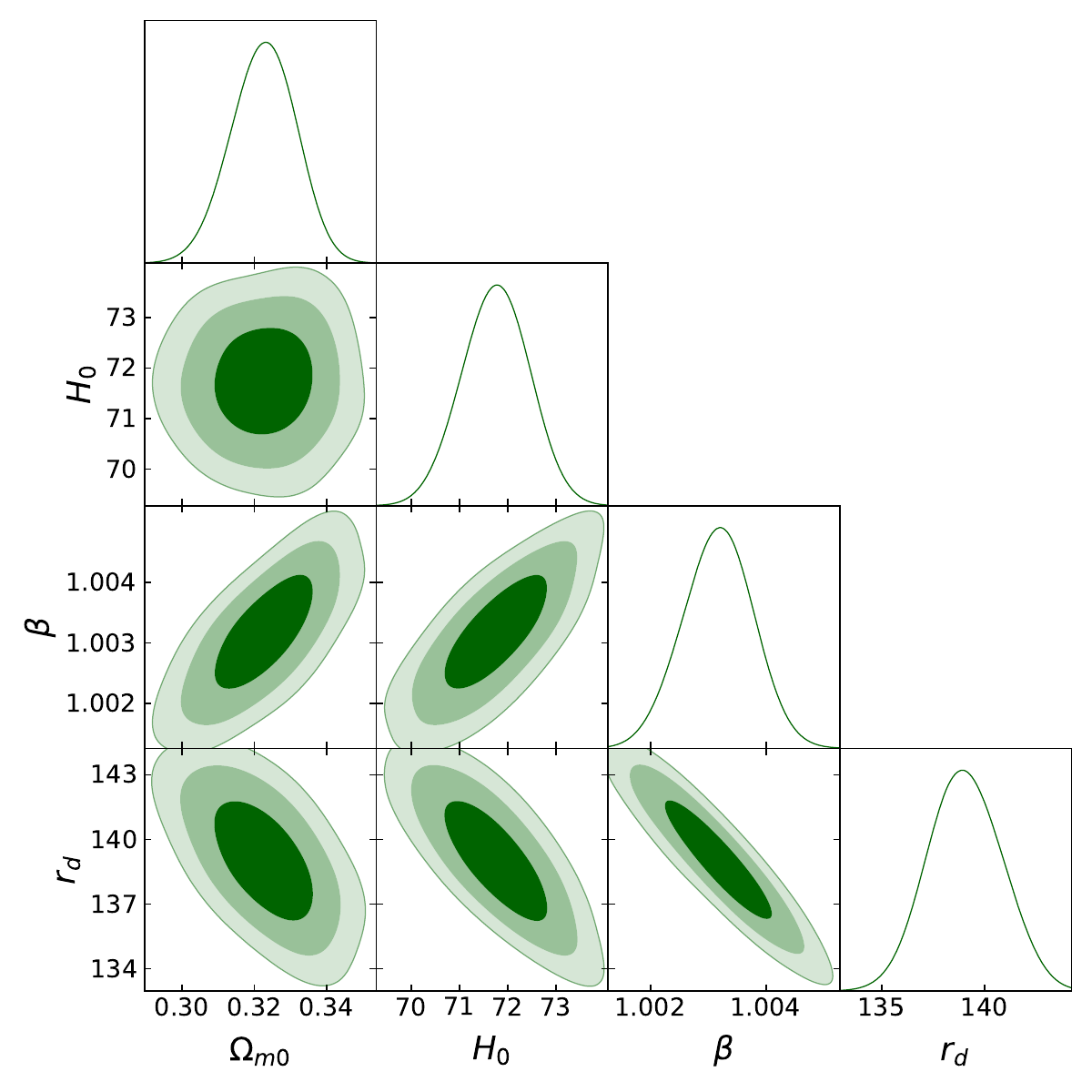}
	\caption{The confidence contours at $1\sigma$, $2\sigma$ and $3\sigma$ levels of the cosmological parameters $\Omega_{m0}$, $H_0$, $\beta$ and $r_d$  in Tsallis cosmology from from combination of DESI BAO+Planck CMB+Pantheon plus datasets.}
	\label{fig:Tc}
\end{figure}
\begin{figure} 
	\centering
	\includegraphics[width=8cm]{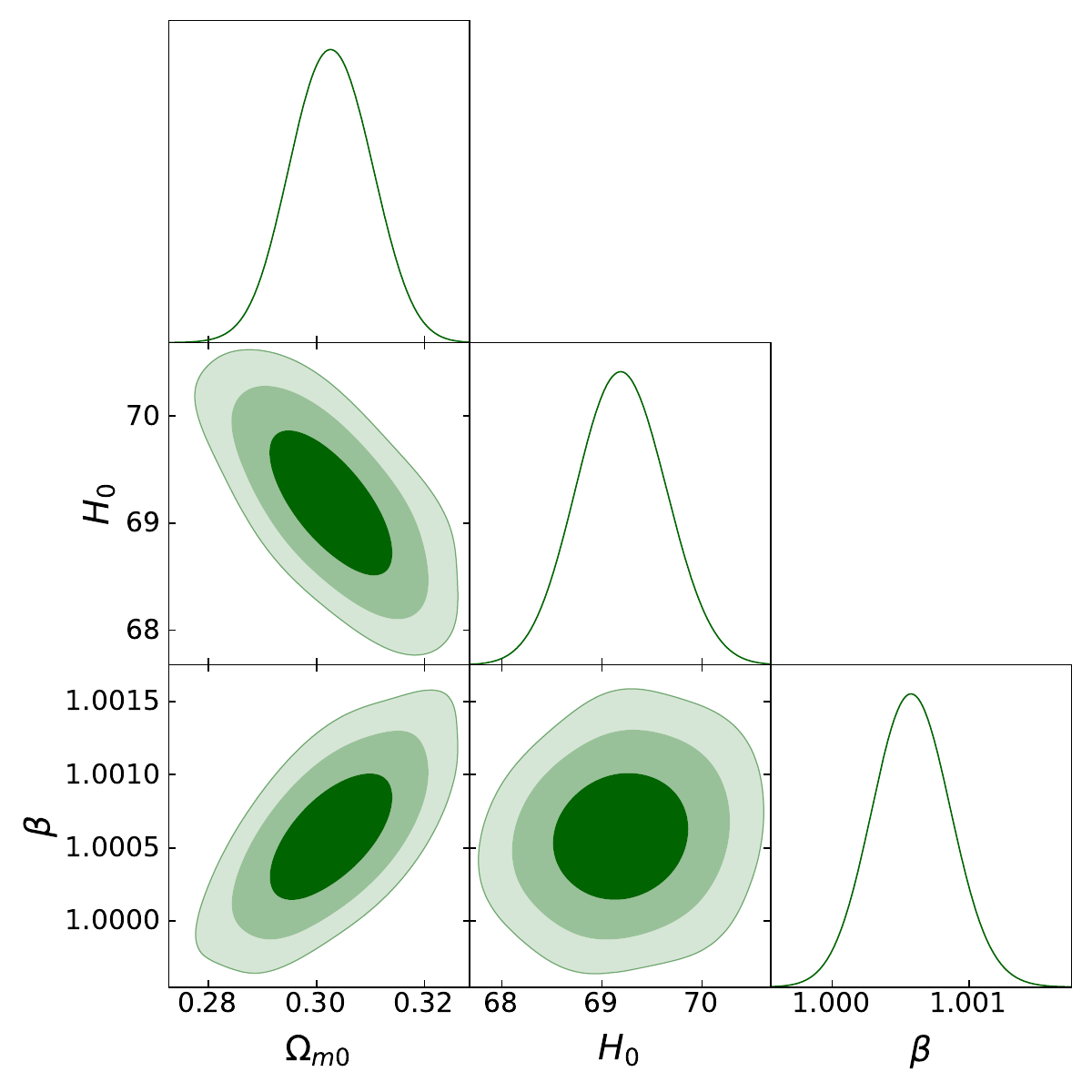}
	\caption{The confidence contours at  $1\sigma$ , $2\sigma$ and $3\sigma$ levels of the cosmological parameters $\Omega_{m0}$ , $H_0$ and $\beta$ in Tsallis cosmology from the combination of DESI BAO+Planck CMB+Pantheon plus datasets, imposing the Planck prior on the sound horizon at baryon drag epoch.}
	\label{fig:TT}
\end{figure}
\subsubsection {DESI BAO + CMB + Pantheon plus}
In this section, we combine all observational datasets in our analysis. Supernovae distance measurements provide supplementary geometric information, helping to constrain the behavior of DE from low redshift observations. In contrast, CMB anisotropy measurements constrain the properties of DE from very high redshift observations. This combination allows us to place cosmological constraints on Tsallis cosmology using observational data spanning all redshifts, from the early to the late-time Universe. For this purpose, we analyze combinations of DESI BAO, Planck CMB anisotropy measurements from 2018, and Pantheon+ datasets. The total chi-square function is given by:
\begin{eqnarray}\label{eq46}
\chi^{2}_{total}=\chi^{2}_{BAO} + \chi^{2}_{CMB} + \chi^{2}_{Pan+}.
\end{eqnarray}
Our numerical results and inferred cosmological parameters are presented in the fifth column of Table \ref{tab:tab1} and Figure \ref{fig:Tc}. We observe that our constraint on the Tsallis parameter $\beta$ deviates from $\beta=1.0$ by more than $3\sigma$, indicating a potential deviation from the standard $\Lambda$CDM cosmology. It is noteworthy that adding Planck CMB anisotropy measurements to our combinations results in more precise constraints with smaller errors compared to previous cases, making it possible to measure the potential deviation from the concordance model.
It is important to note that while we include CMB observations in our analysis, we do not impose the Planck prior on $r_d$, treating this parameter as a free parameter in the DESI BAO measurements. Consequently, the combination of DESI BAO, CMB, and Pantheon+ datasets determines, implying a $4.2\sigma$ deviation from the Planck measurements of $r_d$. Compared to the previous analysis of DESI BAO and Pantheon+ alone, we reduce the deviation of from the Planck value by $1.3\sigma$ due to the inclusion of the CMB dataset while still keeping as a free parameter in the DESI BAO part of the combination.
In the next section, we impose the Planck prior on $r_d$ and examine the results with this preset condition. Previous studies on constraining the $\beta$ value \citep[see][]{Asghari:2021lzu} showed no significant deviation from $\beta=1.0$ when the Planck prior on $r_d$ was imposed and the old SDSS-III \cite{Chuang:2013hya} datasets were used. Additionally, the authors of \citep{Asghari:2021lzu} demonstrated consistency between the $\Lambda$CDM model and Tsallis cosmology by measuring the AIC Criteria values for both models. We find a $5.2\sigma$ deviation between our measurement of $H_0$ and the Planck $\Lambda$CDM value $H_0=67.4\pm 0.5$ km/s/Mpc. This significant tension arises because $r_d$ is treated as a free parameter in our analysis. Finally, we obtain $\Delta \text{AIC} = \text{AIC}_{\Lambda} - \text{AIC}_{T} = 5.8$, indicating moderate evidence against the $\Lambda$CDM cosmology. Hence, the standard $\Lambda$CDM model is less supported by the DESI BAO+CMB+Pantheon Plus data combination than the Tsallis cosmology. Indeed, the AIC criteria analysis aligns with our finding that the deviation of the Tsallis parameter $\beta$ from the $\Lambda$CDM value ($\beta = 1.0$) exceeds the $3\sigma$ confidence region when utilizing the DESI BAO+CMB+Pantheon Plus data combination.
\subsubsection{The impact of Planck prior}
In this section, we impose the Planck prior on $r_d$ in our analysis. We combine the DESI BAO, CMB, and Pantheon+ datasets, where Cepheids calibrate the absolute magnitudes of Pantheon+ SNIa, and the sound horizon at baryon drag epoch in DESI BAO measurements is imposed by the Planck prior $r_d=147.46\pm0.28$ Mpc \cite{Lemos:2023qoy}. Our results are presented in the last column of Table \ref{tab:tab1} and Figure \ref{fig:TT}. 
Our constraints on the Tsallis parameter $\beta$ indicate no significant deviation from $\beta=1.0$ within a $3\sigma$ confidence level. Thus, imposing the Planck prior on $r_d$ in our combined analysis supports the recovery of the standard $\Lambda$CDM cosmology within the framework of the gravity-Tsallis entropy conjecture. Our measurement of the matter's contribution to the Universe's energy budget is consistent with the results from the Planck CMB anisotropy measurements of 2018 \cite{Planck:2018vyg}. 
Furthermore, due to the inclusion of the Planck prior on the BAO measurements, the constraints on $H_0$ show a significant reduction in the discrepancy between our measurement and the Planck $\Lambda$CDM value. Specifically, we now observe a $2.7\sigma$ deviation (less than $3\sigma$) between our measurement and the Planck $\Lambda$CDM value, indicating no significant deviation between our measurement on $H_0$ and Planck-$\Lambda$CDM value. To some degree, this result is similar to what was found in \cite{Basilakos:2023kvk}. Eventually, we obtain $\Delta \text{AIC} = \text{AIC}_{\Lambda} - \text{AIC}_{T} = 2.5$, suggesting weak evidence against the $\Lambda$CDM model. In this case, since we impose the Planck prior on $r_d$, the situation for the $\Lambda$CDM model in the AIC analysis is better compared to the previous case where $r_d$ was a free parameter. Indeed, applying the Planck prior on $r_d$ significantly reduces the deviation of the $\beta$ parameter and also diminishes the evidence against the standard $\Lambda$CDM cosmology.

\section{Conclusion}\label{Sect:6}
The gravity-thermodynamics conjecture is a pivotal concept in modern cosmology, wherein the Einstein field equation is derived from the proportionality of entropy and horizon area, alongside the fundamental thermodynamic relation $\delta Q = T dS$ \cite{Jacobson:1995ab}. This conjecture allows us to derive the standard Friedmann equations in cosmology by applying the Bekenstein-Hawking entropy-area relation.
In the context of the modified Tsallis entropy-area relation \cite{Tsallis:1987eu}, which leads to modified Friedmann equations \cite{Sheykhi:2018dpn}, we explore potential deviations from the standard $\Lambda$CDM cosmology using recent geometrical observations, including DESI BAO \cite{DESI:2024mwx}, Planck CMB anisotropy measurements \cite{Planck:2018vyg}, and the Pantheon+ SNIa compilation \cite{Scolnic:2021amr}.
When analyzing the DESI BAO alone or Pantheon+ dataset alone, our constraints statistically support $\beta=1.0$, indicating no measurable deviations from the standard $\Lambda$CDM cosmology. To tighten constraints on cosmological parameters, we combined datasets. Combining DESI BAO and Pantheon+ datasets initially fully supported the standard $\Lambda$CDM cosmology within the Tsallis entropy framework. However, combining DESI BAO, Planck CMB, and Pantheon+ datasets revealed deviations exceeding $3\sigma$ from the standard $\Lambda$CDM model. This result was achieved without imposing the Planck prior on the sound horizon at the baryon drag epoch, $r_d$, which appeared in DESI BAO analysis. With $r_d$ as a free parameter, we observed significant tensions of $4.2\sigma$ and $5.2\sigma$ between our measurements of $r_d$ and $H_0$ and the Planck-$\Lambda$CDM values, respectively.
By imposing the Planck prior on $r_d$ and analyzing the DESI BAO+CMB+Pantheon+ datasets, we demonstrated support for the standard model within the gravity-thermodynamics conjecture defined by Tsallis entropy. Additionally, the discrepancy between our $H_0$ measurements and the Planck value was reduced to less than $3\sigma$ with the Planck prior on $r_d$.
Finally, we compared the Tsallis and standard $\Lambda$CDM cosmologies from the viewpoint of statistical AIC analysis. We observed that in the cases of DESI BAO alone, Pantheon+ alone, and a combination of both, there is no significant evidence against either model. When we add CMB anisotropic measurements to other data, we observed moderate evidence against the $\Lambda$CDM cosmology. However, imposing the Planck prior on $r_d$ diminishes this evidence. All results obtained in the AIC analysis are in agreement with the results obtained from constraining the $\beta$ parameter.

\section{Acknowledgment}
We would like to extend our gratitude to the anonymous referee for her/his invaluable feedback and constructive comments, which have significantly enhanced the quality of this work.

\bibliographystyle{spphys}
\bibliography{ref}

\label{lastpage}

\end{document}